\documentclass[prl,aps,twocolumn,groupedaddress,showpacs,floatfix]{revtex4}
\usepackage{amsmath,amssymb,multirow,epsfig,bm,subfigure}

\newcommand{\beq}{\begin{equation}}
\newcommand{\eeq}{\end{equation}}
\newcommand{\bea}{\begin{eqnarray}}
\newcommand{\eea}{\end{eqnarray}}

\begin{document}
\title{First-order chiral transition in the compact lattice theory of graphene \\
and the case for improved actions}

\author{Joaqu\'{\i}n E. Drut$^1$, Timo A. L\"ahde$^2$ and Lauri Suoranta$^2$ }
\affiliation{$^1$Department of Physics, The Ohio State University, Columbus, OH 43210--1117, USA}
\affiliation{$^2$Helsinki Institute of Physics and Department of Applied Physics, 
Aalto University, FI-02150 Espoo, Finland}

\begin{abstract}
A comparison of the compact and non-compact lattice versions of the low-energy theory of graphene is 
presented. The compact theory is found to exhibit a chiral phase transition which appears to be of
first order, at a critical coupling of $\beta_c^{} = 0.42 \pm 0.01$. We confirm that the non-compact
theory exhibits a second-order transition at $\beta_c^{} = 0.072 \pm 0.003$, and determine the effects
of UV-divergent tadpole contributions in both cases. Upon tadpole improvement of the
non-compact theory we find $\beta_c^\text{TI} = 0.163 \pm 0.002$, which strengthens the case for a 
semimetal-insulator transition in graphene at strong Coulomb coupling.
Finally, we highlight the need for systematic studies using improved lattice actions.
\end{abstract}

\date{\today}

\maketitle

Graphene, a sheet of $sp^2$-bonded carbon, has become an attractive 
candidate for nanoscale electronics due to its many remarkable and desirable 
properties~\cite{GeimNovoselov,CastroNetoetal}. These include 
high carrier mobility at room temperature, great mechanical and tensile strength, in addition to 
chemical stability and impermeability. While the hexagonal lattice symmetry suggests that graphene is
a semimetallic material exhibiting massless Dirac quasiparticles, 
the possibility that graphene might become an excitonic insulator due to strong Coulomb 
interactions has recently been revived~\cite{CastroNetoPauling}.

In a series of papers within the Lattice Monte Carlo~(LMC) 
framework~\cite{DrutLahde1,DrutLahde2+3}, the semimetal-insulator transition, which manifests 
itself as the spontaneous breaking of a $U(4)$ chiral symmetry, was found to happen at a critical 
coupling of $\alpha_c^{} \sim 1.1$, which is intermediate between that of graphene on a SiO$_2^{}$
substrate ($\alpha \sim 0.8$) and suspended graphene ($\alpha \sim 2.1$). 
LMC studies of a closely related Thirring-like model, including 
the determination of the renormalized Fermi velocity, were reported in 
Ref.~\cite{HandsStrouthos}. These results belong to a larger class of LMC studies, 
such as those of Quantum Electrodynamics in $2+1$~(QED$_3^{}$)~\cite{QED3_lattice}
and $3+1$~(QED$_4^{}$)~\cite{QED4_lattice} dimensions, and of four-fermion theories such as the 
Thirring~\cite{Thirring_lattice} and Gross-Neveu~\cite{GrossNeveu_lattice} models.

Spontaneous chiral symmetry breaking in LMC is typically studied using staggered 
fermions~\cite{Staggered}, as chiral symmetry is then partially preserved at finite lattice spacing 
$a$. In practical LMC simulations the continuum limit is recovered in the vicinity 
of second-order phase transitions, where the relevant correlation lengths $\xi$ diverge. 
In the case of graphene, such a continuum description is attainable on the critical line of vanishing bare mass 
and (inverse) coupling $\beta \geq \beta_c^{} \sim 0.072$, with $\alpha_c^{} \equiv 1/(4\pi\beta_c^{})$.
In the strong coupling 
phase $\beta < \beta_c^{}$ the Coulomb interaction induces the formation of particle-hole pairs, with 
a binding energy that yields an intrinsic cutoff scale $\Lambda \sim a^{-1}$, such that the 
limit $a \to 0$ is not well defined (or rather, it defines an unstable theory). 
Keeping $a$ finite, however, one may approach the critical line and recover a well-defined theory, 
with bound states at strong coupling and massless fermions at weak coupling.

The requirement of exact gauge invariance on the lattice necessitates the use of ``gauge links" 
$U \equiv \exp(i\theta)$, with $\theta$ the lattice gauge field. Gauge links have the side effect of 
introducing vertices of higher order in $a$, such as photon-photon interactions,
which are absent in the continuum theory. Such vertices also yield potentially large
``tadpole" contributions, where the naive power-counting in $a$ is cancelled by UV divergences.
Consequently, Lattice QCD simulations employ various levels of ``improvement"~\cite{ImprovedActions}, 
in order to minimize the impact of such discretization artifacts.
We report here a systematic study of the tadpole effects in the graphene theory, which were first pointed 
out in Ref.~\cite{Giedt}, and their impact on the determination of $\beta_c^{}$. 
We also present a comparison of the non-compact and compact lattice theories of graphene, where the latter
case involves photon self-interactions which are absent in the former. 

The Euclidean action of the lattice theory of graphene is conventionally split into gauge and
fermion components, such that $S_E^{} = S^g_E + S^f_E$. In the compact formulation, the gauge field
$\theta$ enters into $S^g_E$ in terms of link variables, giving
\beq
S^{g,c}_E[\theta] = \beta \sum_{\bf n} {\left [3 - \sum^3_{i=1} \Re 
\left(U_{\bf n}^{} U^\dagger_{{\bf n}+{\bf e}_i^{}}\right) \right]},
\label{SgaugeC}
\eeq
where $\Re(x)$ denotes the real part of $x$, 
$\beta \equiv 1/g^2$ is the inverse coupling, $\bf n$ a site on the (3+1)-dimensional space-time 
lattice, and ${\bf e}_\mu^{}$ a unit vector in the direction $\mu$. 

Early on, the analogous compact formulation of QED$_4^{}$ was found to exhibit a first-order 
transition~\cite{QED4_compact} where the breaking of chiral symmetry is coincident with the condensation 
of magnetic monopoles. While the addition of a four-Fermi interaction to QED$_4^{}$ makes
it possible to isolate the chiral symmetry breaking transition, the resulting theory may 
belong to a different universality class than continuum QED~\cite{QED4_fermi}. 
In order to avoid this situation for the case of 
graphene, an attractive option is the non-compact formulation
\begin{eqnarray}
S^{g,nc}_E[\theta] &=& \frac{\beta}{2} \sum_n
{\left[\sum^3_{i=1} \left(\theta_{{\bf n}}^{} - \theta_{{\bf n} + {\bf e}_i^{}}^{}\right)^2_{} \right]},
\label{Sg}
\end{eqnarray}
which is free of photon self-interactions, and  
is known to support a second-order transition which may be identified with the spontaneous breaking of
chiral symmetry in the continuum theory. Thus, Eq.~(\ref{Sg}) has become the standard choice 
for LMC studies of Abelian gauge theories, such as QED$_3^{}$~\cite{QED3_lattice}, 
QED$_4^{}$~\cite{QED4_lattice} and we have used it for graphene. The first objective of this study is to
characterize the compact version of the low-energy theory of graphene, which differs significantly 
from QED$_4^{}$ as the spatial gauge links are constant and the fermions propagate in two spatial dimensions
only.

The fermion action $S_E^f$ is identical in the compact and non-compact theories, and in the staggered
fermion formulation it is given by
\begin{eqnarray}
S^f_E[\bar{\chi},\chi,\theta] &=& 
-\sum_{{\bf n},{\bf n}'}
\bar\chi_{\bf n}^{} \: D_{{\bf n},{\bf n}'}^{}[\theta] \: \chi_{{\bf n}'}^{},
\label{Sf}
\end{eqnarray}
where $({\bf n},{\bf n}')$ denote the sites of a (2+1)-dimensional space-time sublattice, and
gauge invariance is retained by coupling the staggered spinors $\chi_{\bf n}^{}$ to the gauge field 
via link variables in the time direction. The staggered Dirac operator is given by 
(see e.g. Ref.~\cite{Rothe})
\begin{eqnarray}
D_{{\bf n},{\bf n}'}^{}[\theta] \!\!&=&\!\!
\frac{1}{2} \left[\delta_{{\bf n}+{\bf e}_0^{},{\bf n}'}^{}\,U_{{\bf n}}^{} 
- \delta_{{\bf n}- {\bf e}_0^{},n'}^{}\,U_{{\bf n}'}^{\dagger}\right]
\label{Df} \\
&& +\:\frac{v}{2} \sum_i \eta_{i,{\bf n}}^{} 
\left[\delta_{{\bf n}+{\bf e}_i^{},{\bf n}'}^{} - \delta_{{\bf n}-{\bf e}_i^{},{\bf n}'}^{}\right] 
+ m_0^{}\,\delta_{{\bf n},{\bf n}'}^{}, \nonumber 
\end{eqnarray}
where $\eta_{1,{\bf n}}^{} = (-1)^{n_0^{}}$ and $\eta_{2,{\bf n}}^{} = (-1)^{n_0^{} + n_1^{}}$. The Fermi
velocity $v$ can be absorbed into the remaining parameters, giving $\beta \equiv v/g^2$ and 
$m \equiv m_0^{}/v$. Simulations are thus conventionally performed in terms of $(\beta,m)$ and $v = 1$.
The mass term acts as a symmetry breaking parameter, without which the chiral condensate $\sigma$
would be zero at finite volume.
The limit $m \to 0$ is reached by extrapolation, using an equation of state~(EOS) of the form
$m = f(\sigma,\beta)$, which describes a second-order transition with critical exponents
$\delta$ and $\bar\beta$. A detailed description of the EOS can be found in Ref.~\cite{DrutLahde2+3}.

\begin{figure}[t] 
\epsfig{file=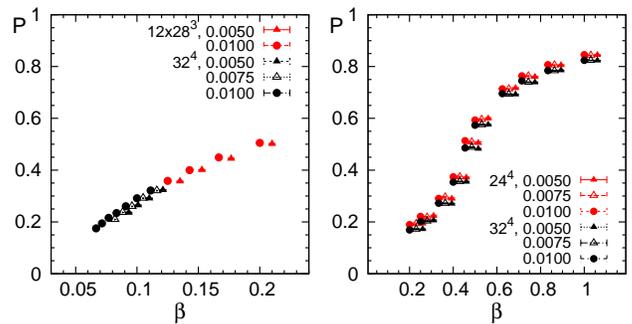, width=\columnwidth}
\caption{(Color online) Plaquettes $\langle P \rangle$ as a function of $\beta$
for the non-compact (left) and compact (right) theories, for Eq.~(\ref{u0def}) averaged over the 
fermionic sublattice. Our notation is $N_z^{} \times N_x^3$, where $N_x^{}$ is the extent of the fermion 
sublattice and $N_z^{}$ that of the bulk dimension. Black datapoints are for a $32^4$ lattice, while red 
(gray) points are for $12\times 28^3$. To illustrate the depencence on $m$, the datapoints have been 
shifted horizontally relative to $m = 0.010$, and for the compact case the $24^4$ data have also been 
shifted vertically relative to the $32^4$ data.}
\label{Fig:Plaquette}
\end{figure}

``Tadpole improvement" (TI) is a non-perturbative method due to Lepage and Mackenzie~\cite{Lepage} 
that accounts for the UV divergent tadpole contributions by renormalizing the link field $U$. The effect of the 
UV modes is encoded in the function $u_0^{}$, which depends on the input parameters of the simulation 
and is to be determined {\it a posteriori}. A conventional definition of $u_0^{}$ is
\beq
u_0^{} \equiv \langle P \rangle^{1/2}_{}, \quad
P = \frac{1}{V} \sum_{{\bf n}} U_{\bf n}^{} U^\dagger_{{\bf n}+{\bf e}_i^{}},
\label{u0def}
\eeq
in terms of the plaquette $P$. It should be noted that the power $1/2$ (instead of $1/4$ as in Lattice QCD)
is due to the smaller number of fluctuating gauge links.

Recently, Ref.~\cite{Giedt} has reported that $u_0^{}(\beta,m)$ deviates significantly
from unity, and thus the effects of TI are likely to be significant. Therefore, the second objective 
of this study is to determine how TI affects the determination of $\beta_c^{}$ and the 
critical exponents. In the non-compact theory, modifications due to TI are restricted 
to the fermion action, whereas in the compact case the gauge action is affected as well. 

As in Ref.~\cite{Giedt}, we consider the plaquettes in the ($x,t$) and ($y,t$) planes, however unlike 
Ref.~\cite{Giedt} we compute $\langle P \rangle$ by summing over the plaquettes in the (2+1)-dimensional 
fermionic sublattice only. This choice is appropriate for the non-compact theory, and the resulting numerical 
differences are insignificant for the subsequent analysis. Our results for $P$ are shown 
in Fig.~\ref{Fig:Plaquette}. We find that $P$ is independent of $m$ up to statistical fluctuations, and
thus we define $u_0^{}(\beta)$ as the average of $P$ over $m$.

In practice, TI amounts to replacing the link field $U$ in Eqs.~(\ref{SgaugeC}) and~(\ref{Sf}) 
according to $U \to U/u_0^{}$, 
where $u_0^{}$ is to be determined {\it a posteriori} using Eq.~(\ref{u0def}). If the staggered spinors 
are rescaled as $\chi \equiv \sqrt{u^{}_0}\,\chi'$, one can define an ``improved" Dirac operator
\begin{eqnarray}
D^I_{{\bf n},{\bf n}'}[\theta] \!\!&=&\!\!
\frac{1}{2} \left[\delta_{{\bf n}+{\bf e}_0^{},{\bf n}'}^{}\,U_{{\bf n}}^{} 
- \delta_{{\bf n}- {\bf e}_0^{},{\bf n}'}^{}\,U_{{\bf n}'}^{\dagger}\right]
\label{DfI} \\
&& + \:
\frac{v'}{2} \sum_i \eta_{i,{\bf n}}^{} 
\left[\delta_{{\bf n}+{\bf e}_i^{},{\bf n}'}^{} - \delta_{{\bf n}-{\bf e}_i^{},{\bf n}'}^{}\right] 
+ m'_0\,\delta_{{\bf n},{\bf n}'}^{}, \nonumber 
\end{eqnarray}
where
\beq
\sigma' \equiv \sigma / u_0^{}, \quad
v' \equiv u_0^{} v, \quad 
m'_0 \equiv u_0^{} m^{}_0,
\label{presc}
\eeq
such that the tadpole-improved observables may be computed using the original unimproved gauge
configurations, provided that the input parameters of the simulation and the spinors are
reinterpreted according to Eq.~(\ref{presc}).
As $u_0^{} < 1$, the net effect of TI (apart from possible shifts of the
critical coupling $\beta_c^{}$) is to make the condensate smaller in the spontaneously broken phase,
an effect which increases with decreasing $\beta$. For the compact gauge action
in Eq.~(\ref{SgaugeC}), TI leads to a similar prescription $g' \equiv u_0^{} g$, while
the non-compact action in Eq.~(\ref{Sg}) is not directly modified as it involves no gauge links.

\begin{figure}[t] 
\epsfig{file=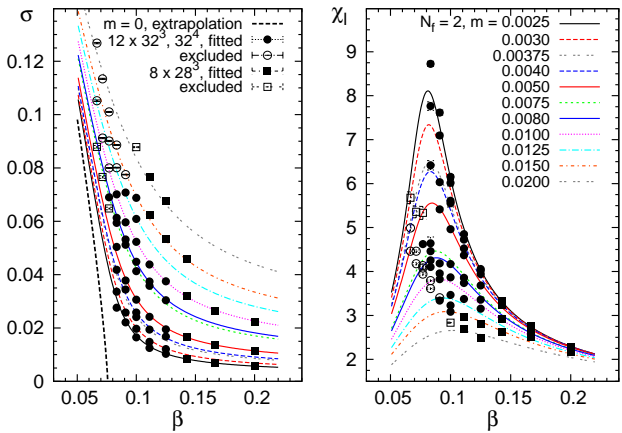, width=\columnwidth}
\caption{(Color online) Chiral condensate $\sigma$ (left) and susceptibility $\chi_l^{} \equiv 
\partial\sigma/\partial m$ (right) in the non-compact theory, together with an EOS fit to the unimproved 
lattice data. The fitted parameters are $\beta_c^{} = 0.072 \pm 0.003$ and $\delta = 2.3 \pm 0.3$, with 
$\bar\beta \simeq 1$. The data are obtained for lattices of size $12\times 32^3$ and $32^4$ (circles) and 
$12\times 28^4$ (squares). Errors for the individual datapoints were obtained with the standard 
block-jackknife method. Note the deviations from the EOS (scaling violations) at small $\beta$ and 
large $m$.}
\label{Fig:CondensateChi_NC}
\end{figure}

\begin{figure}[t] 
\epsfig{file=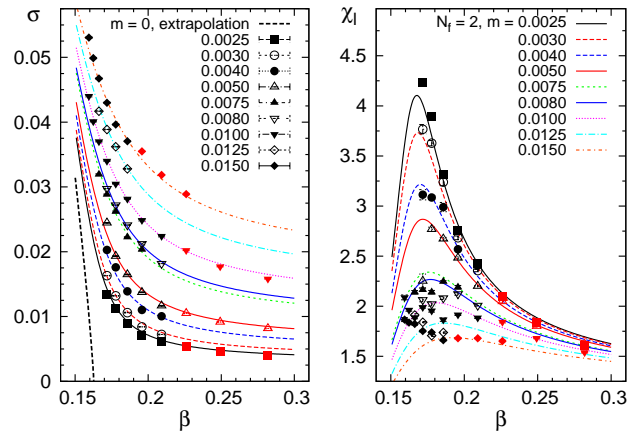, width=\columnwidth}
\caption{(Color online) Chiral condensate $\sigma$ (left) and susceptiblity $\chi_l^{}$ (right) in the 
non-compact theory after TI, together with an EOS fit. Note the significantly reduced scaling
violations compared to the unimproved data. The optimal fit is compatible with $\bar\beta = 1$, giving
$\beta_c^{\text{TI}} = 0.163 \pm 0.002$ and $\delta = 2.2 \pm 0.1$. The errors are mostly systematical, due 
to finite-volume effects and residual scaling violations. The black symbols denote data on $12\times 32^3$ 
and $32^4$ lattices, while red (gray) symbols denote data for $12\times 28^3$.}
\label{Fig:CondensateChi_NCI}
\end{figure}

\begin{figure}[b] 
\epsfig{file=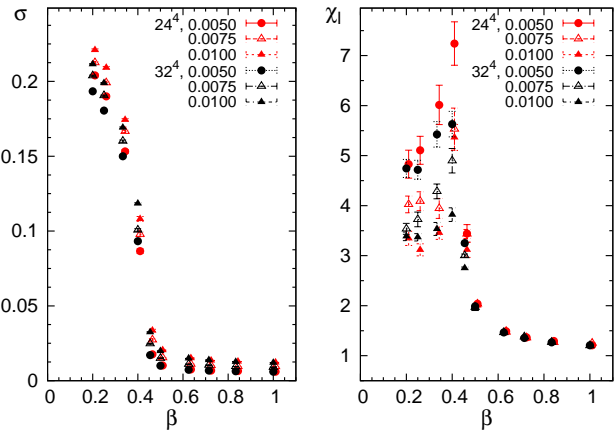, width=\columnwidth}
\caption{(Color online) Chiral condensate $\sigma$ (left) and susceptibility $\chi_l^{}$ (right) for the 
compact, unimproved theory as a function of $\beta$. A very sharp, possibly first-order 
transition is observed at $\beta_c^{} = 0.42 \pm 0.01$. Data on a $24^4$ lattice are shown in
red (gray) and are for clarity slightly offset from the (black) datapoints on a $32^4$ lattice. 
The volume dependence of
$\chi_l^{}$ is greatly enhanced in the spontaneously broken phase.}
\label{Fig:CondensateChi_C}
\end{figure}

\begin{figure}[t] 
\epsfig{file=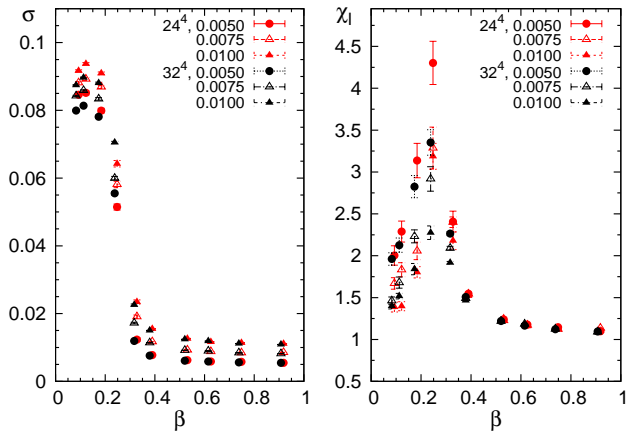, width=\columnwidth}
\caption{(Color online) Chiral condensate $\sigma$ and susceptibility $\chi_l^{}$ in the 
compact theory of graphene after TI. The presentation is similar to Fig.~\ref{Fig:CondensateChi_C}. 
The critical coupling $\beta_c^\text{TI} = 0.27 \pm 0.01$ has been shifted closer to that of the 
non-compact theory. At $\beta_c^{}$, the data indicate a discontinuous jump of $\Delta_\sigma^{} \sim 0.04$ 
in the condensate.
\label{Fig:CondensateChi_CI}}
\end{figure}

The primed parameters in Eq.~(\ref{presc}) represent the input for the LMC calculation. Once $u^{}_0(\beta)$ 
has been mapped out, the tadpole-improved (unprimed) quantities can be determined. 
Apart from the rescaling of the chiral condensate according to $\sigma' \to u_0^{}\sigma$, we find for
the compact case
\beq
\beta \equiv \frac{v}{{g}^2} = 
\frac{v'/u_0^{}}{g'^2/u_0^2} = u_0^{} \beta',
\label{c_presc}
\eeq
while for the non-compact case we have
\beq
\beta \equiv \frac{v}{{g}^2} = 
\frac{v'/u^{}_0}{g'^2} = \frac{\beta'}{u_0^{}},
\label{nc_presc}
\eeq
and we note that $m \equiv m_0^{}/v$ remains unmodified. Both of these results
differ from the prescription $\beta \equiv u^2_0 \beta'$
which was applied in Ref.~\cite{Giedt}. However, such a choice would be valid for the compact theory in
the absence of TI for the fermion action. It is noteworthy that the non-compact theory
would have $\beta' \equiv \beta$ if the tadpoles of fermionic origin were ignored.
Our results for the non-compact theory are shown in 
Figs.~\ref{Fig:CondensateChi_NC} and~\ref{Fig:CondensateChi_NCI} for the unimproved and TI cases, respectively.
Similarly, the results for the compact theory are given in Figs.~\ref{Fig:CondensateChi_C} 
and~\ref{Fig:CondensateChi_CI}. Most of our data for the
non-compact case are taken from Ref.~\cite{DrutLahde2+3}, except for the data on $32^4$ lattices close to 
the critical point.

As the tadpole correction $u_0^{}(\beta) \sim 0.5$ close to the chiral phase transition in the 
non-compact theory, the effects of TI can potentially be dramatic. We find that the agreement with the 
EOS improves and the effects of outlying data points are lessened, which is due to decreased scaling 
violations at small $\beta$. While the net effect on the critical exponents is slight (these remain 
compatible with the values of Ref.~\cite{DrutLahde2+3}), the change in $\beta_c^{}$ is larger as a 
consequence of Eq.~(\ref{nc_presc}). After TI, we find $\beta_c^\text{TI} \simeq 0.16$, which is a factor 
of $\sim 2$ larger than the unimproved value. On the other hand, the compact theory exhibits an abrupt 
transition close to $\beta \sim 0.45$, and is furthermore not compatible with the EOS description of 
Ref.~\cite{DrutLahde2+3}, although TI does bring $\beta_c^{}$ closer to the non-compact result, as 
evidenced by Fig.~\ref{Fig:CondensateChi_CI}. As a first estimate, we find $\Delta_\sigma^{} \simeq 0.04 \pm 0.01$ 
for the jump in the condensate.

It is intriguing that $\beta_c^\text{TI} > \beta_c^{}$ in the non-compact theory, as it suggests that graphene 
samples on a SiO$_2^{}$ substrate may become insulating as well. However, LMC studies of the Fermi velocity in
graphene~\cite{HandsStrouthos} indicate a significant downward renormalization of $v$ due to
strong Coulomb interactions, thereby again decreasing the physical value of $\beta_c^{}$. Further LMC
studies of the electrical conduction properties of graphene are in progress~\cite{DrutLahdeCond}.

In conclusion, we find that the effect of TI is not as drastic as reported in 
Ref.~\cite{Giedt}, particulary in the non-compact theory, where $\beta_c^{}$ is shifted to a larger 
(instead of a smaller) value. The predicted semimetal-insulator transition in suspended graphene is 
therefore unlikely to be an artifact of tadpole effects. 
While the full chiral symmetry is not realized in LMC simulations with staggered fermions at small $\beta$
(see Refs.~\cite{QED4_lattice,Giedt}), it is known to eventually be restored in the vicinity of the
tricritical point $(\beta = \beta_c^{}, m = 0)$. However, this conceptual weakness may be
inconsequential, as the accuracy of the extrapolation can be systematically improved by obtaining additional
data closer to the tricritical point. Nevertheless, simulations with overlap fermions~\cite{Overlap} 
may serve to further clarify this point, and to extend the scope of the present work.


\begin{acknowledgments}

We acknowledge support under DOE Grant No.~DE-FC02-07ER41457 (UNEDF SciDAC Collaboration), 
and NSF Grant No.~PHY--0653312.
This work was supported in part by an allocation of computing time from the Ohio Supercomputer Center. 
We thank Richard Furnstahl, Junko Shigemitsu, Heechang Na, Simon Hands, Costas Strouthos, Ari Harju and 
Joel Giedt for instructive discussions 
and comments.

\end{acknowledgments}



\end{document}